\documentclass[journal]{IEEEtran}
\ifCLASSINFOpdf
\else
   \usepackage[dvips]{graphicx}
   \usepackage{subfigure}
   \graphicspath{{../eps/}}
   \DeclareGraphicsExtensions{.eps}
\fi
\usepackage{amsmath}
\usepackage{empheq}
\usepackage{upgreek}

\begin{document}
%
\title{Non-Invasive Imaging Method of Microwave Near \\Field Based on Solid State Quantum Sensing}
%
%
%

\author{Bo~Yang,~Guanxiang~Du${^{*}}$,~Yue~Dong,~Guoquan~Liu,~Zhenzhong~Hu,~and~Yongjin Wang
\thanks{Manuscript received Sep 15, 2017; revised Jan 4, 2018.}
\thanks{Bo Yang, Guanxiang Du, Yue Dong, Zhenzhong Hu and Yongjin Wang are with College of Telecommunications \& Information Engineering, Nanjing University of Posts and Telecommunications, No.66 Xin Mofan Road, Nanjing, China. Bo Yang works at JiangSu Institute Of Quality and
Standardization. (${^{*}}$correspondence e-mail: duguanxiang@njupt.edu.cn).}
\thanks{Guoquan Liu is an assistant professor of college of pharmacy, Peking University, Beijing, College Road 38, China.}}

%
%

\markboth{Submitted to IEEE T-MTT, January~4~2018}%
{Shell \MakeLowercase{\textit{et al.}}: Bare Demo of IEEEtran.cls for IEEE Journals}
%



\maketitle

\begin{abstract}
In this paper, we propose a non-invasive imaging method of microwave near field using a diamond containing nitrogen-vacancy centers. We applied synchronous pulsed sequence combined with charge coupled device camera to measure the amplitude of the microwave magnetic field. A full reconstruction formulation of the local field vector, including the amplitude and phase, is developed by measuring both left and right circular polarizations along the four nitrogen-vacancy axes. Compared to the raster scanning approach, the two dimensional imaging method is promising for application to circuit failure analysis. A diamond film with micrometer thinness enables high-resolution near field imaging. The proposed method is expected to have applications in monolithic-microwave-integrated circuit chip local diagnosis, antenna characterization, and field mode imaging of microwave cavities and waveguides.
\end{abstract}


%
\IEEEpeerreviewmaketitle

\section{Introduction}
%
%
%
%
\IEEEPARstart{M}{icrowave} near-field imaging technology plays an important role in material testing, monolithic microwave integrated circuit (MMIC) design, semiconductor device design \cite{Berweger2015Microwave}, nondestructive detection, and medical diagnosis \cite{De2017A},\cite{Mcguinness2011Quantum}. Traditional microwave (MW) near field probes employ metallic open ended waveguides (OEW) to sense the local microwave field. The OEW probe is typically placed at distances not smaller than several wavelengths from the antenna under test (AUT), which brings unavoidable coupling between the AUT and the probe \cite{Vinetti2008A}. Moreover, the commercially available probe dimension is on the order of millimeter scale. In addition, the field mapping efficiency of probe technique is rather low because of the requirement of doing raster scan over the AUT \cite{Sch2014Development}. Because of these shortcomings, traditional technology has been unable to meet the requirements of MW chip near field imaging, antenna near field characterization and medicine applications.

In recent years, quantum atomic systems, such as the alkali vapor cell and the diamond nitrogen-vacancy (NV) center, have been proposed for imaging \cite{Cui2013Quantum} and sensing of MW near fields \cite{Sedlacek2013Atom},\cite{Cao2013Spintronic}.
The resolution of vapor cell technique, limited by the diffusion of vapor atoms \cite{Horsley2014Imaging}, is usually on the order of 100 $\upmu$m. The diamond NV center, on the other hand, provides a much higher nanoscale resolution \cite{Maurer2010Far},\cite{Appel2015Nanoscale} by using a 10$-$20 nm diamond nanocrystal containing a single NV center \cite{Wang2015High}, or a bulk diamond sample together with a charge coupled device (CCD) camera \cite{J2008High},\cite{Tamburello2015Magnetic}. Chipaux group developed a magnetic imaging method based on a 250 $\upmu$m diamond thin slab \cite{Tamburello2015Magnetic},\cite{Shao2016Wide}. Horsley reported a type of frequency-tunable MW imaging system based on an atomic vapor cell \cite{Horsley2016Frequency}. From previous work, we can see great potential exists for MW imaging using the diamond probe, because its non-invasive nature, nanoscale resolution, and high sensitivity. Moreover, diamond thin film with micrometer thickness can be placed at the proximity of MW integrated circuits under test.

\begin{figure}
  \centering
  \includegraphics[width=0.8\linewidth]{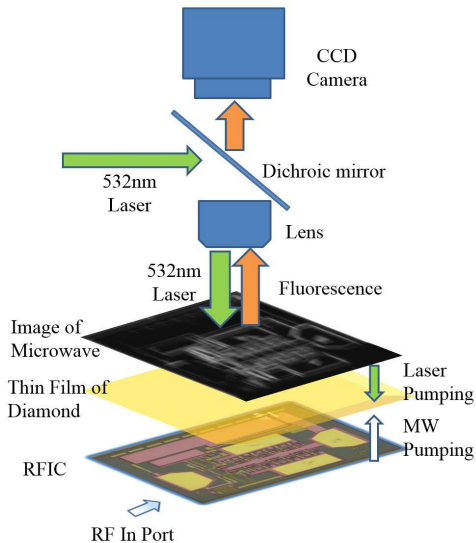}\\
  \caption{Schematic diagram of direct imaging mode applied in the field of microwave chip imaging. A diamond thin film is placed on the microwave chip surface, and a green laser is projected on the diamond thin film. Scattered fluorescence captured by the CCD camera form the microwave chip imaging. The camera uses this image to subtract the reference image without microwaves and obtains the image of the pure microwave field.}\label{Fig1}
\end{figure}

We herein bring forward a novel frequency-tunable and high-efficiency MW near-field imaging method based on a CCD camera. The method provides a spatial resolution that approaches the optical diffraction limit. As indicated in Fig.\ref{Fig1}, our group developed a non-confocal (NCFM) optical experimental system based on Kohler illumination \cite{Sohn2006Koehler}. The system not only supports raster scanning mode, but also supports direct imaging mode based on diamond thin film. In the field of view (FOV) of Kohler illumination, the optical and static magnetic field is homogeneous. A CCD camera with an optical fluorescence (FL) filter is used to acquire a series of FL images of the diamond thin film. Each pixel of the CCD camera corresponds to a physical area in the diamond film. Use of the NV ensemble enhances the signal to noise ratio by a factor of $\sqrt N$, where N is the total number of NV centers in the area \cite{Tamburello2015Magnetic}.

In quantum field, the diamond nitrogen vacancy (NV) center is found to be sensitive to magnetic field and microwave (MW) fields \cite{Glenn2017Micrometer}. NV center is a unique quantum system that combines solid state quantum bits (qubits) with coherent optical transitions. It consists of six electrons: four electrons are valence electrons; two electrons are free electrons. The lattice structure of the NV center has a C$_{3V}$ symmetry structure, and its quantum energy level is composed of the $^3$A$_2$ spin triplet structure ground state: a 3E structure excited state, and 1A structure meta-stable state. There are two paths of radiative transition and non-radiative transition from the excited state to the ground state. The spin state ${m_s}= \pm1$ of the NV center would prefer non-radiative transition. In contrast, ${m_s}=0$ would prefer radiative transition. The MW field can change the distribution of qubits in NV center. Therefore, the fluorescence intensity reflects the change of MW field.

The proposed method employs pulsed MW to manipulate qubits of the NV center. The change of microwave pulse length leads to a periodic flip of the distribution ratio between ${m_s}=0$ and ${m_s}=\pm1$ at the Rabi frequency \cite{Popa2006Pulsed}. Here, ${{\rm{\Omega }}_{mw}} = 2\pi {\gamma _{nv}}\left| {{B_ \bot }} \right|$, ${{\rm{\Omega }}_{mw}}$ represents the Rabi frequency of the NV center, ${\gamma _{nv}}$ denotes the gyromagnetic ratio of the NV center, and ${\gamma _{nv}} = 28\ k{\rm{Hz}}/\mu {\rm{T}}$. $\left|{{B_ \bot }}\right|$ is the amplitude of the circular polarized MW perpendicular to the NV axis. So the Rabi frequency can be used to measure the absolute value of the intensity of the microwave field.

This work implements MW field imaging for the first time. To do so, a multi-channel pulse generator is used to synchronize CCD, laser switch and MW switch. A differential measurement method between reference frame and capture frame is used to reduce the measurement noise. Each frame integrates N repeated sequence to increase signal to noise ratio. We developed a software scan MW imaging method, which reconstruct the MW imaging by calculating the strength the MW strength for each pixel based on ODMR spectrum or Rabi frequency. A full vector MW reconstruction formulation of the local field vector, including the amplitude and phase, is developed, which is deduced from both left and right circular polarizations along the four NV axes that can be measured. The whole experimental system has been verified by measuring a miniature spiral antenna. Compared with HFSS simulation, the spatial resolution of new system has been remarkably improved. So the near field distribution image of microwave quantum can reflect the actual microwave near field distribution of the real antenna more accurately. This technique makes it possible for researcher to visual design microwave antennas and devices \cite{Yang1998High}.


\section{EXPERIMENTAL SETUP AND METHODS}
\subsection{Experimental System Construction}
\begin{figure}
  \centering
  \includegraphics[width=0.9\linewidth]{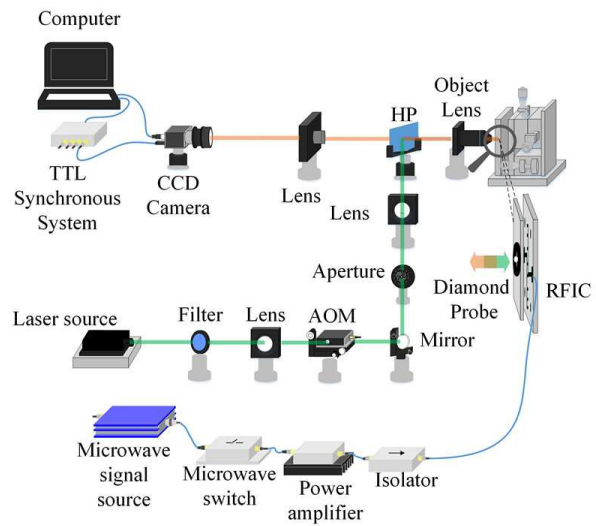}\\
  \caption{Schematic diagram of experimental system. The TTL synchronous system take the responsible of synchronizing CCD Camera, MW signal source, AOM and MW switch, and the line of synchronous system is ignored for concise.}\label{Fig2}
\end{figure}

The experimental system consists of a MW subsystem and an optical subsystem. Fig.\ref{Fig2} displays the structure of the experimental system. The MW source supports an external hardware trigger, the frequency resolution is 0.01 Hz, and the maximum output radio frequency (RF) power is +14 dBm. This system utilizes the MW pulse to manipulate the quantum state of two energy-level systems. A MW switch is connected with a MW source in series. The typical rise and fall time is 5 ns, and the minimum length of the MW pulse is 12 ns, which is limited by the pulse generator with a time resolution of 2 ns. A type of mini-circuits ZHL-16W-43-S+ power amplifier is used before the AUT; RFIC can be seen as a special AUT. The typical gain is 45 dB. In order to protect the amplifier, a microwave isolator absorbs the reflected waves from the AUT .

The optical subsystem uses a 532 nm green laser to excite the diamond NV centers, and fluorescence of the NV center of the diamond is collected by a lens and the CCD camera. The laser output power is 300 mW.  Owing to the spectral impurity of the laser source, we use a 532 nm optical line filter to clean up the laser line. In order to convert continuous light into pulsed light, the laser is focused on the crystal of an acousto-optical modulation device (AOM).  The rise time of AOM is 25 nm, and the pulse length of the laser in the pulse mode is 300 ns. An aperture is used to select the first-order diffraction beam of AOM, and an additional lens is implemented to collimate the diverged beam out of the AOM. Moreover, a dichroic mirror reflects the laser to the objective lens, which focuses the parallel light into the diamond sample. An electronic controlled three-dimensional translation stage of Beijing Zolixis is used to move the diamond sample into the field of view. In direct imaging mode, to realize wide field of view, a third lens is placed in front of the dichroic mirror. The focus point of the third lens should be aligned with the back focus point of the object lens, where the illuminated area is uniformly enlarged. This is called Kohler illumination \cite{Sohn2006Koehler}. The fluorescence of the NV center is collected by the objective lens with a 0.54 numerical aperture, the focal length is 2.5 cm. The high numerical aperture object lens enhances the efficiency of fluorescence collection. Long working focal length ensures sufficient operating space.
\begin{figure}[!t]
\centering
\subfigure[] {\includegraphics[width=0.2\textwidth]{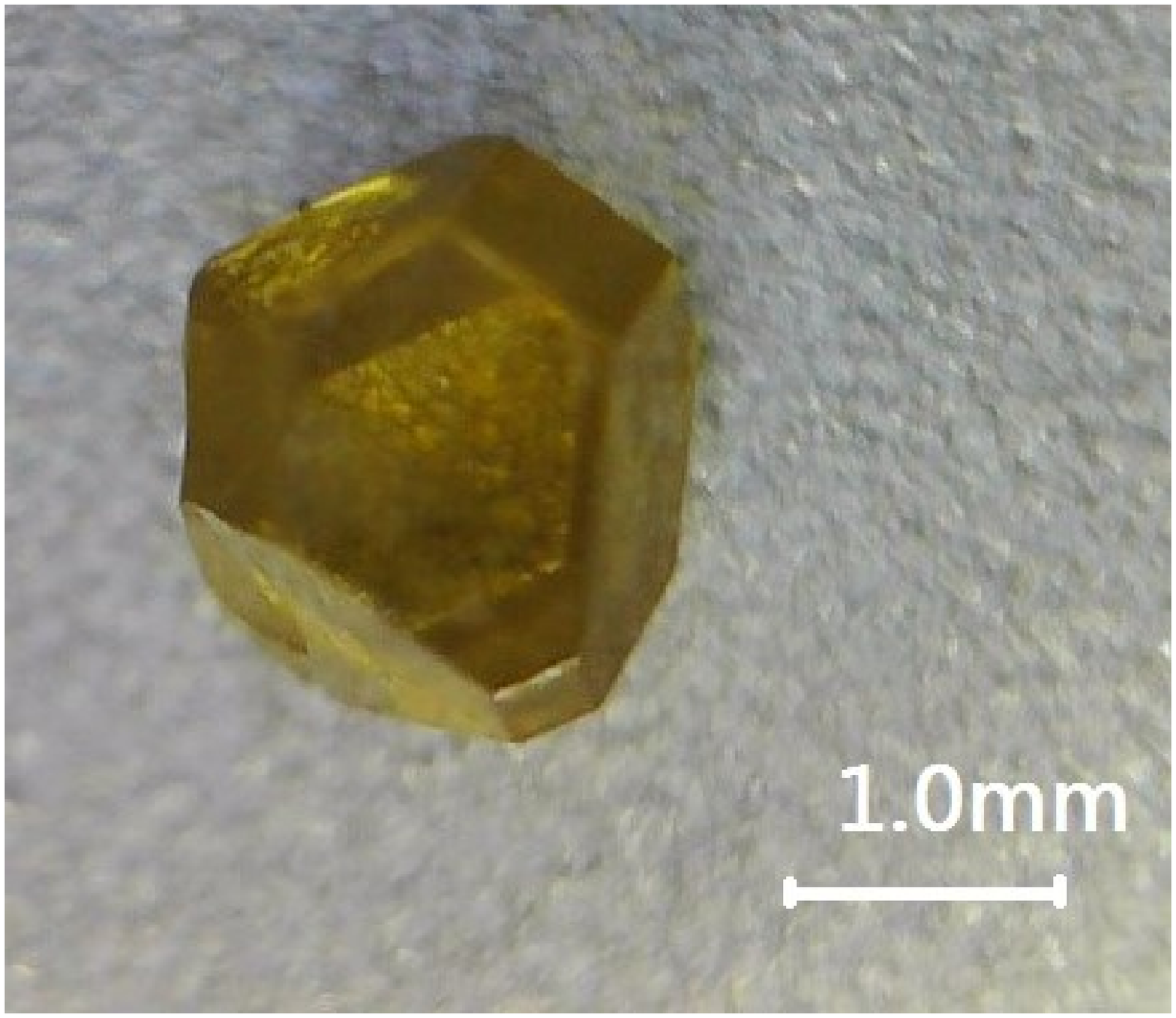}}
\subfigure[] {\includegraphics[width=0.2\textwidth]{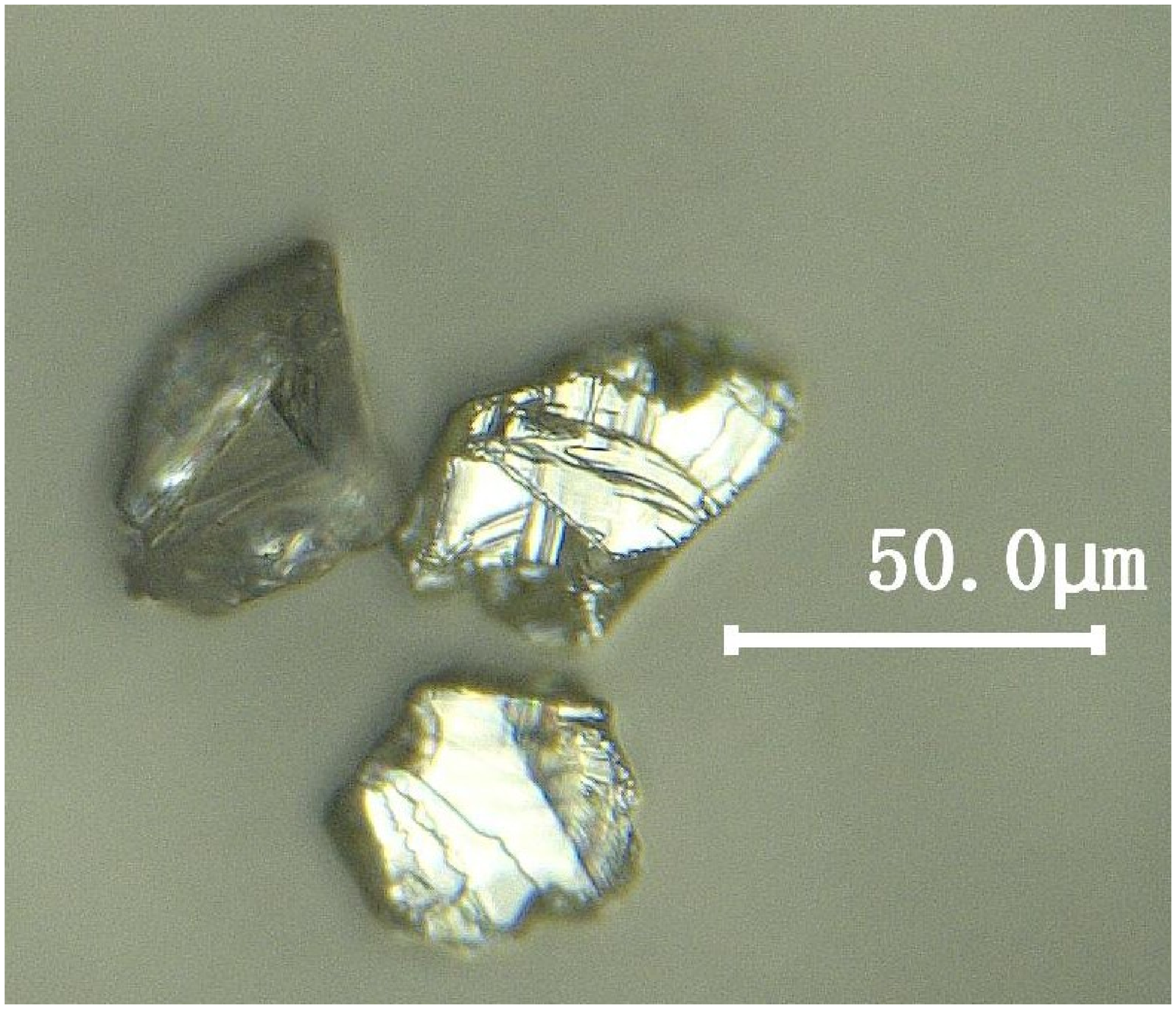}}
\caption{(a) Microscope picture of a 2 mm diamond sample. The photograph was obtained using a 10X microscope. The length of the reference ruler is 1 mm. (b) Microscope picture is the photo of a 60 $\upmu$m diamond. The image was obtained by using a 400X microscope. The length of the reference ruler is 50 $\upmu$m.}
\label{Fig3}
\end{figure}

\begin{figure}
  \centering
  \includegraphics[width=0.9\linewidth]{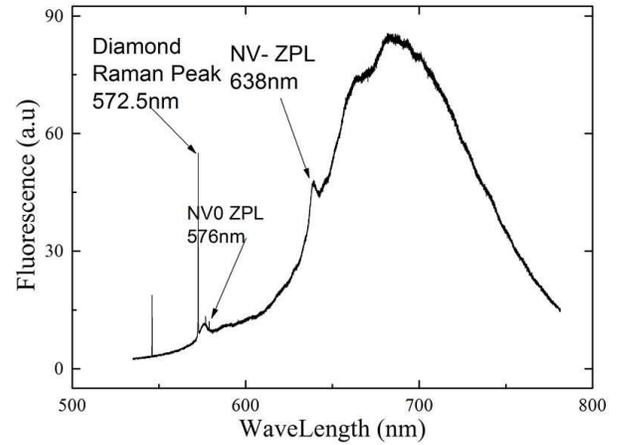}\\
  \caption{Diagram showing the Raman spectrum of a 60 $\upmu$m diamond sample. The diamond Raman feature peak is 572.5 nm, NV$^-$ ZPL and NV$^0$ ZPL is on the right of the diamond Raman peak. The FL spectrum of the diamond is between 600 and 800 nm.}\label{Fig4}
\end{figure}

In recent years, several groups have developed NV based magnetometers and MW imaging instruments by using a single NV center in a bulk diamond \cite{Appel2015Nanoscale},\cite{Maze2008Nanoscale}. The fundamental shot noise in the measurement of single electron spin constrains the potential sensitivity of the optical magnetometers. In this work, we choose the bulk diamond as the MW probe, thus sacrificing the spatial resolution for the benefit of easy realization. The diameters of the candidate diamond are 250 nm, 1 $\upmu$m, 10 $\upmu$m, 50 $\upmu$m, 60 $\upmu$m and 2 mm, and the composition and origin of the diamond particles of different diameters are the same. Fig.\ref{Fig3}(a) shows that the color of the 2 mm diamond is yellow. the 60 $\upmu$m, 50 $\upmu$m, and 10 $\upmu$m diamond appears light green as Fig.\ref{Fig3}(b). The 1 $\upmu$m diamond presents as white powder, and the 250 nm diamond is an agglomerate powder with a brown color. The diamond sample we selected is the Ib type diamond. For the micron-scale diamond sample, we use the Raman \cite{Manson1990Raman} spectrometer to analyze the optical spectrum of the NV center, as shown in Fig.\ref{Fig4}. The 576 nm zero-phonon line (ZPL) indicates the existence of NV$^0$, and the 638 nm ZPL indicates the existence of NV$^-$ \cite{Zheng2010Study}. Thus, we use a 600 nm high pass filter to filter the fluorescence of NV$^-$.  The FL contrast is the absolute value of the difference between the FL intensity of the MW and FL intensity of No MW.

To optimize the laser power for maximum FL contrast, we researched the optical property of the candidate diamond and finally selected the 60 $\upmu$m diamond as the MW probe for better performance with respect to the laser pumping efficiency and FL contrast. We determined that the FL contrast firstly increased with the laser power. Then, the FL contrast reached the maximum value. We referred to this laser power corresponding to the maximum FL contrast working point as the laser power threshold. When the laser power exceeded the threshold, the FL contrast rapidly decreased.

With the size of the diamond becoming smaller, an increasing amount of laser power is needed to make diamond particles work at the maximum FL contrast working point. This principle is always valid for a diamond dimension below 60 $\upmu$m. Thus, we selected the pumping laser power that is slightly smaller than the threshold. However, the diameter of the diamond would not be large without a limitation. The 2 mm diameter diamond does not present a better FL efficiency than the 60 $\upmu$m diamond because the fluorescence is refracted to the other side of the diamond. The surface of the 60 $\upmu$m diamond is rougher than that of the 2 mm diamond, and the rough surface diamond can more easily emit fluorescence than a single crystal diamond. Therefore, we strived to make the surface of the diamond rougher for easier detection of FL.
\subsection{Measurement and Imaging Method}
Based on the experimental system of our group, a new quantum MW visual design and diagnosis method was implemented on the platform. The procedure of this method included four steps: optical focusing, frequency tuning, FL data collecting, and imaging reconstruction.

The optical focusing procedure establishes the mapping relationship between the spatial location of the NV color centers and the pixel points. The Kohler illumination projects the pumping laser on the diamond. The AUT should be moved into the focus of the laser. The diamond thin film is deposited on the AUT as close as possible. When a clear image of the diamond thin film appears in the lens view, the optical focusing is finished.

Frequency tuning is based on Zeeman splitting technology. We impose a uniformly external static magnetic field on the diamond probe, and the direction of the external static magnetic field should be parallel to the axial direction of the NV color center. The optically detected magnetic resonance (ODMR) technology provides a method to detect qubits in the NV center. As shown in Fig. 5, the diamond MW resonance points split into eight resonance peaks under the action of the external static magnetic field. The eight resonance peaks are symmetrically distributed on the two sides of the zero-field splitting (ZFS). At room temperature, the ZFS frequency is 2.87 GHz. Under the influence of temperature and stress, the ZFS will deviate from 2.87 GHz. The resonance frequency larger than ZFS corresponds to the energy difference between the spin state ${m_s}=0$ and ${m_s}=+1$; otherwise, it corresponds to the energy difference between ${m_s}=0$ and ${m_s}= -1$. There are four possible NV axes in the diamond crystals. Thus, in total, there are 8 resonance peaks in ODMR spectrum. The distance between resonance frequency and ZFS is decided by the strength of external static magnetic field. Thus, the strength of the external static magnetic field should be adjusted to a certain value so that the resonance frequency is the same as the working frequency of AUT. As shown in Fig. 5, the NV center under the resonance between ${m_s}=0$ and ${m_s}= +1$ is sensitive to left-hand circular polarization MW. We can calculate the left-hand circular polarization magnetic field strength of MW through the FL intensity; the algorithm is discussed further below.
\begin{figure}
  \centering
  \includegraphics[width=0.9\linewidth]{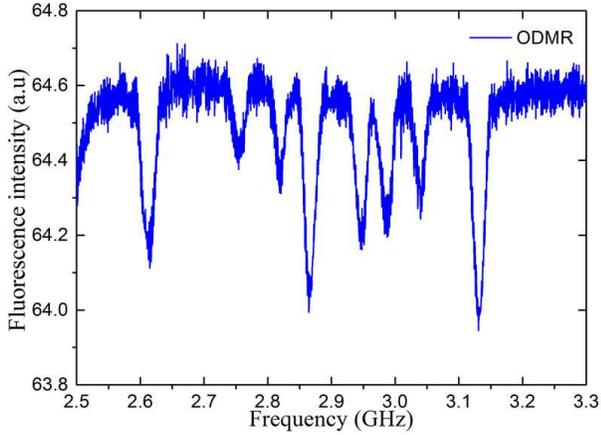}\\
  \caption{ODMR spectrum of 60 $\upmu$m diamond measured in the experimental system. There are eight resonance points in the ODMR spectrum, which are symmetrically distributed on the left and right sides of zero-field splitting (ZPS). The distance of a pair of symmetrical peaks relates to the static magnetic projection on the NV axis. The height of each peak relates to the strength of the MW field perpendicular to the NV axis.}\label{Fig5}
\end{figure}
Because the total MW magnitude information is composed of left-hand and right-hand circular polarization MW, obtaining only the left-hand polarized information is not adequate. Therefore, we must additionally tune the NV center to the electron spin resonance state between ${m_s}=0$ and ${m_s}= -1$. If we hold the strength and reverse the direction of the external static magnetic field, the Zeeman splitting of transition energy between ${m_s}=0$ and ${m_s}= +1$ will be reversed to ${m_s}= -1$. Accordingly, we can measure the magnetic field strength of the right-hand circular polarization MW.

\begin{figure}
  \centering
  \includegraphics[width=0.9\linewidth]{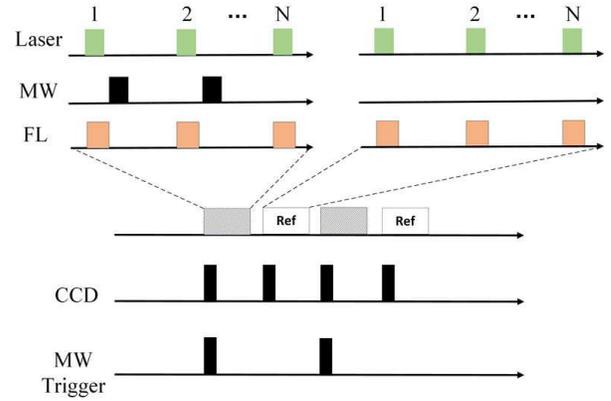}\\
  \caption{Typical control pulse sequence of the experimental system. N laser pulses, N MW pulses, and N FL pulses form a frame. The Ref frame does not contain MW pulses. CCD pulses control the camera to obtain a photo for each frame. MW trigger pulses are MW frequency step pulse signals, which change the MW frequency step to the next point every two frames. In this example, the number N equals 1E-4, the pulse width of the laser is 300 ns, the pulse width of the MW is 50 ns, and the period of laser pulses is 3 $\upmu$s. For controlling the pulse synchronous sequence of Rabi, MW is fixed to the resonance frequency, and the length of MW pulse is linearly increased by a 2 ns step. For the control pulse synchronous sequence of Ramsey \cite{J2008High}, there are two $\pi {\rm{/2}}$ MW pulses with intervals linearly increasing by 2ns steps.}\label{Fig6}
\end{figure}

FL data collecting is used to gather and integrate the FL pulse information of the NV center by using the high-speed CCD camera. Before FL data collecting, the synchronous pulse sequence should be designed to ensure that the laser pulse, MW switch pulse, MW frequency source external trigger, and CCD camera external trigger (four parts) synchronously work together. The control pulse synchronous sequences are depicted in Fig.\ref{Fig6}. The control pulse signal is composed of a laser pulse signal, MW pulse signal, camera trigger signal, and MW step trigger signal. All synchronous sequences are generated from the TTL signal generator, which supports four standalone signal paths. From Fig.\ref{Fig6}, the MW sources step into next frequency point every two CCD frames : the first frame is an image of the FL intensity integration with MW resonance; the second is an image of the FL intensity integration without MW resonance. The difference between the first and second frames is the pure FL contrast information of the MW resonance, and the second frame is the reference frame.
The integration of N fluorescence pulses divided by N reaches closer to the average value of FL. ODMR can roughly reflect the information of the external static magnetic field and MW field strength. Compared with ODMR, the Rabi frequency measurement provided an absolutely accurate MW measurement method. We also designed a Ramsey oscillation experiment to accurately measure the electron spin decay time T2 \cite{J2008High}. The synchronous sequence of Rabi and Ramsey is described in Fig.\ref{Fig6}.

\begin{figure}
  \centering
  \includegraphics[width=0.9\linewidth]{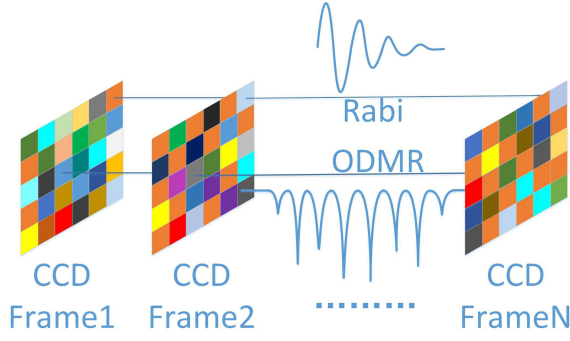}\\
  \caption{Schematic diagram of ODMR/Rabi software scan imaging method. CCD camera captures N frames pictures along time axis. The control sequence refers to Fig.\ref{Fig6}.}\label{Fig7}
\end{figure}

Here, we propose a software scan MW imaging method that calculates the ODMR spectrum or Rabi frequency for each pixel. By using the ODMR/Rabi synchronization sequence of Fig.\ref{Fig6}, the CCD camera acquires a series of FL images. As shown in Fig.\ref{Fig7}, the FL image is separated into many units. All images of the same position unit in each frame consist of a time signal sequence. By selecting Rabi or ODMR control sequences, we can calculate the Rabi frequency or produce the ODMR spectrum for each unit, and then compute the MW strength of each unit. The size of the unit represents the imaging resolution. We use the number M of pixels of each unit to describe the resolution. If M increases, the resolution decreases. However, each unit will include more NV centers; thus, the signal-to-noise ratio will increase. Therefore, M represent the performance of whole imaging system. The empirical value in experimental system is 22. Obviously, the minimum limit of M value is 1.

\subsection{MW Field Reconstruction Algorithm}
MW reconstruction algorithm is used to reconstruct the image of the MW field by using multiple frames of CCD fluorescent images. The algorithm is based on the Rabi frequency or area of the ODMR spectrum to calculate the corresponding MW magnetic field strength. Firstly, we used the right-handed circularly polarized MW as an example. Rabi frequency has a fixed mathematical relationship with the MW magnetic field strength. We calculate the Rabi frequency according to the data fitting formula \eqref{E-2}.
\begin{equation}\label{E-2}
\begin{split}
&F\left( {{\gamma _{NV}},{B_{ - MW}},\tau } \right) =
 \\ {F_b} + {F_c}\cos (&2\pi{\rm{ }}{\gamma _{NV}}\left| {{B_{ - MW}}} \right|\left( {\tau  + d} \right)){e^{\frac{{ - \left( {\tau  + d} \right)}}{{{\tau _R}}}}}
\end{split}
\end{equation}

In formula \eqref{E-2}, ${F_b}$ is the fluorescence in the $\left| 0 \right\rangle $ state, ${{B_{ - MW}}}$ is right-hand circular polarization MW magnetic field strength, ${F_c}$ is the Rabi oscillation amplitude and ${\tau _R}$ is the decay time of the Rabi oscillation, $d$ is the relative delay of system, and ${{\gamma _{NV}}}$ is the gyromagnetic ratio of NV. The mathematical expressions of Ramsey oscillation are shown in the formula \eqref{E-3}.

\begin{equation}\label{E-3}
\begin{split}
&F\left( {{f_{lo}},{B_{NV}},\tau } \right) =\\
 {F_b} + {F_c} \cos &\left( {2\pi \left( {{f_{lo}} - {f_{NV}}} \right)\left( {\tau  + d} \right)} \right){e^{\frac{{ - \left( {\tau  + d} \right)}}{{{T_2}}}}}
\end{split}
\end{equation}

In formula \eqref{E-3}, ${F_b}$ is the fluorescence in the $\left| 0 \right\rangle $ state, ${F_c}$ is the contrast of fluorescence, and ${{T_2}}$ is the decay time of electron spin. ${{f_{NV}}}$ is the resonance frequency of certain NV axial, ${f_{NV}} = 2.87GHz \pm \frac{{g{u_b}}}{h}{B_{NV}},\frac{{g{u_b}}}{h} = 28\;M{\rm{Hz}}/m{\rm{T}}$, the external MW frequency. Therefore, although the intensity of the external MW field can be accurately measured with the Rabi frequency formula, the measurement of ${{B_{ - MW}}}$ based on the Rabi frequency is limited by the purity of the NV color center in the diamond on account of the shorter ${\tau _R}$ time.

Compared with the Rabi frequency, the ODMR curve presents a Lorentzian line shape around each resonance point. Thus, we use the Lorentzian function to fit ODMR data in a small frequency range (Fig.\ref{Fig5}). The fitting function refers to formula \eqref{E-4}.

\begin{equation}\label{E-4}
\begin{split}
S\left( {C,\Gamma } \right) = {N_0}\left( {1 - \frac{{C {\Gamma ^2}}}{{{\Gamma ^2} + {\gamma ^2}{{\left( {B - {B_0}} \right)}^2}}}} \right)
\end{split}
\end{equation}
In \eqref{E-4}, ${N_0}$ is the unperturbed FL intensity in regions of the ODMR spectrum far detuned  from resonance, and $C$ is the difference of FL intensity between the resonant and unperturbed regions. $\Gamma$ denotes the full width at half maximum (FWHM). Thus, the ${C\Gamma}$ approximation represents the strength of the external MW field ${{B_{ - MW}}}$ under the conditions of constant laser power.

To date, we can measure the right-handed circularly polarized MW intensity by two methods, Rabi and ODMR. Similarly, the left-handed circularly polarizing MW intensity can be obtained by reversing the direction of the static magnetic field, retaining the magnetic field strength as the same as
the previous. In fact, people usually describe an electromagnetic wave of one point in the XY plane by two orthogonal components, the mathematical expressions of the MW to be measured in the XY plane are given as \eqref{E-5}. When $\varphi$ equals %
$n\pi$, the MW is the linear polarization MW. When $\varphi$ equals $\left( {n + 1} \right)\pi/2$, the MW is the circular polarization MW. When $\varphi$ equals another value, the MW is the elliptical polarization MW.
\begin{equation}\label{E-5}
\begin{split}
{B_{MW}} &= {B_x}sin\left( {\omega t} \right) + i{B_y}sin\left( {\omega t + \varphi } \right)\\
 &= {B_{ + MW}} + {B_{ - MW}}
\end{split}
\end{equation}
\begin{equation}\label{E-6}
\begin{split}
{B_{MW}} = {e^{i\omega t}}\left( {\frac{{{B_x}}}{{2i}} + \frac{{{B_y}{e^{i\varphi }}}}{2}} \right) + {e^{ - i\omega t}}\left( {\frac{{ - {B_x}}}{{2i}} + \frac{{ - {B_y}{e^{ - i\varphi }}}}{2}} \right)
\end{split}
\end{equation}
\begin{equation}\label{E-7}
\begin{split}
\left\{ {\begin{array}{*{20}{c}}
{{B_{{\rm{ + }}MW}} = \frac{{{B_y}\cos \varphi }}{2} + i\left( {\frac{{{B_y}\sin \varphi  - {B_x}}}{2}} \right)}\\
{{B_{ - MW}} = \frac{{ - {B_y}\cos \varphi }}{2} + i\left( {\frac{{{B_y}\sin \varphi  + {B_x}}}{2}} \right)}
\end{array}} \right.
\end{split}
\end{equation}
\begin{equation}\label{E-8}
\begin{split}
\left\{ {\begin{array}{*{20}{c}}
{\left| {{B_{ + MW}}} \right| = \sqrt {{{\left( {\frac{{{B_y}}}{2}} \right)}^2} + {{\left( {\frac{{{B_x}}}{2}} \right)}^2} - \frac{{{B_y}{B_x}sin\varphi }}{2}} }\\
{\left| {{B_{ - MW}}} \right| = \sqrt {{{\left( {\frac{{{B_y}}}{2}} \right)}^2} + {{\left( {\frac{{{B_x}}}{2}} \right)}^2} + \frac{{{B_y}{B_x}sin\varphi }}{2}} }
\end{array}} \right.
\end{split}
\end{equation}
\begin{equation}\label{E-9}
\begin{split}
&\mathop {{B_{MW}}}\limits^.  \\
= \mathop {{e_x}}\limits^. {B_x}sin\left( {\omega t} \right) + \mathop {{e_y}}\limits^. {B_y}&sin\left( {\omega t + \varphi } \right) + \mathop {{e_z}}\limits^. {B_z}sin\left( {\omega t + \psi } \right)
\end{split}
\end{equation}
As we can see from formula \eqref{E-5} to \eqref{E-8}, only ${{B_{ + MW}}}$ and ${{B_{ - MW}}}$ are not enough to derive ${B_x}$ and ${B_y}$ due to the unknown variable $\varphi$.
Previous calculation methods usually assume a precondition of the linear polarization MW and circular polarization MW, and then replace ${B_x}$ and ${B_y}$ with ${{B_{ + MW}}}$ and ${{B_{ - MW}}}$. However, these existing testing methods are not applicable for the elliptical polarization MW. Because $\varphi$ is an additional unknown variable for elliptical polarization MW, we must define more equations to solve the variable $\varphi$.

We assume a three dimensional XYZ space is based on the reference coordinate system as $X,Y,Z$. We assume the crystal coordinate system of a diamond crystal as ${X_1},{Y_1},{Z_1}$, and a ${100}$ crystal orientation corresponds to the ${X_1}$ axis, ${010}$ crystal orientation corresponds to the ${Y_1}$ axis, ${001}$ crystal orientation corresponds to the ${Z_1}$ axis. We use $\theta ,\rho ,\beta$ to define the relationship between crystal coordinate system ${X_1},{Y_1},{Z_1}$ and reference coordinate system $X,Y,Z$, and $\theta ,\rho ,\beta$ represent azimuthal angle, pitching angle and roll angle. We define ${e_x},{e_y},{e_z}$ as a unit vector of the ${X}$ ${Y}$ ${Z}$ reference coordinate system. Thus, we can use formula \eqref{E-9} to describe the actual three dimensional MW vector.
Matrix $C1\left( {\theta ,\rho ,\beta } \right)$, $C2\left( {\theta ,\rho ,\beta } \right)$, $C3\left( {\theta ,\rho ,\beta } \right)$, $C4\left( {\theta ,\rho ,\beta } \right)$  can be used to describe the projection of the MW vector on the vertical plane of the NV axis 1, NV axis 2, NV axis 3, NV axis 4.
\begin{figure}
  \centering
  \includegraphics[width=0.9\linewidth]{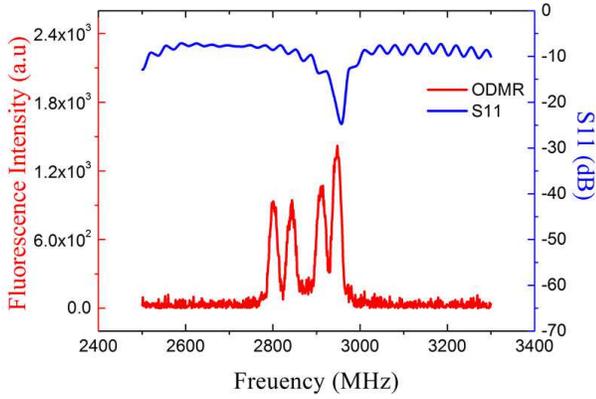}\\
  \caption{A typical resonance frequency of ODMR is tuned to the working frequency of antenna. The testing frequency is 2951 MHz, the resonance peak located into the passband of antenna. The 3 dB passband of antenna is from 2928.7MHz to 2991.6MHz.}\label{Fig8}
\end{figure}
We can thus define the projection of MW on axis 1 as ${B_{MW - 1}} = {B_{MW}}C1\left( {\theta ,\rho ,\beta } \right)$; the microwave should be expressed as a vector matrix. The definition of ${B_{MW - 2}}$, ${B_{MW - 3}}$, ${B_{MW - 4}}$ is similar with ${B_{MW - 1}}$. Finally, we measure the left-hand and right-hand polarizations MW intensity of the four NV axis. Combined with \eqref{E-5} and \eqref{E-8}, we can define eight equations of ${B_{MW + n}}$ and ${B_{MW - n}}$ to associate with ${B_x},{B_y},{B_z},\varphi ,\psi ,\theta ,\rho ,\beta $.
\begin{figure}
  \centering
  \includegraphics[width=0.9\linewidth]{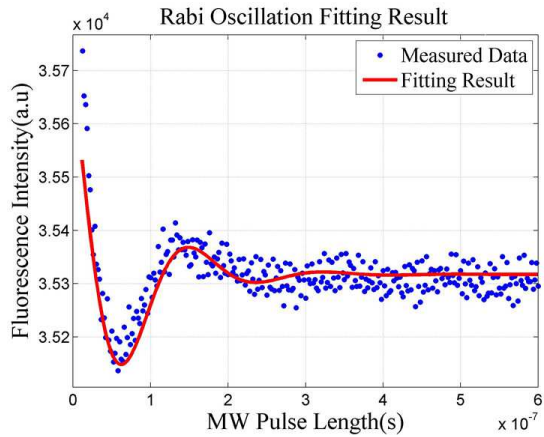}\\
  \caption{A typical Rabi frequency curve of diamond NV center.  The data fitting result is ${\gamma _{nv}} = 28kHz/\mu {\rm{T}},{\tau _R} = 71ns,{B_{ - MW}} = 2.057 \times {10^2}\mu T$.}\label{Fig9}
\end{figure}
The 8 variables correspond to 8 equations, so the equations are complete. Since the equation solving process is very difficult, we must utilize Matlab to obtain the solution of the equations. The solution of equations not only gives the intensity of electromagnetic field in the reference coordinate system, but it also gives the relationship between reference coordinate system and crystal coordinate system. This has never been achieved in previous studies.

\section{EXPERIMENTAL RESULT}
In order to verify the feasibility of the experimental system, we designed an experiment to measure the MW strength of the near field. A miniature spiral antenna was tested. Before testing, the resonance frequency was tuned to the working frequency of the antenna. Fig.\ref{Fig8} describes the tuning process. In this example, the antenna working frequency is 2951 MHz.

\begin{figure}
  \centering
  \includegraphics[width=0.82\linewidth]{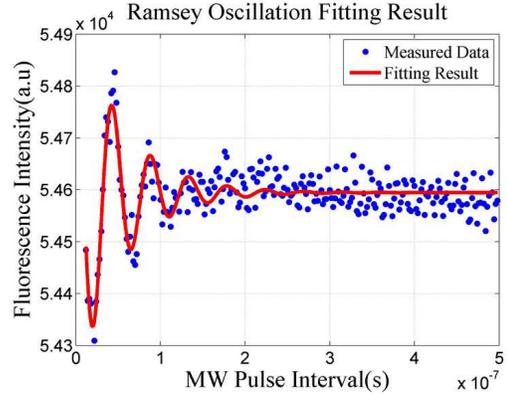}\\
  \caption{A typical Ramsey frequency curve of diamond NV center. The data fitting result is ${f_{NV}} = 2922.1MH{\rm{z}}$ , ${f_{LO}} = 2900MHz$ , $\frac{{g{\mu _b}}}{h} = 28MHz/\mu T$, ${T_2} = 52ns$, ${B_{NV}} = 1.857mT$.}\label{Fig10}
\end{figure}

\begin{figure}[!t]
\centering
\subfigure[] {\includegraphics[width=0.24\textwidth]{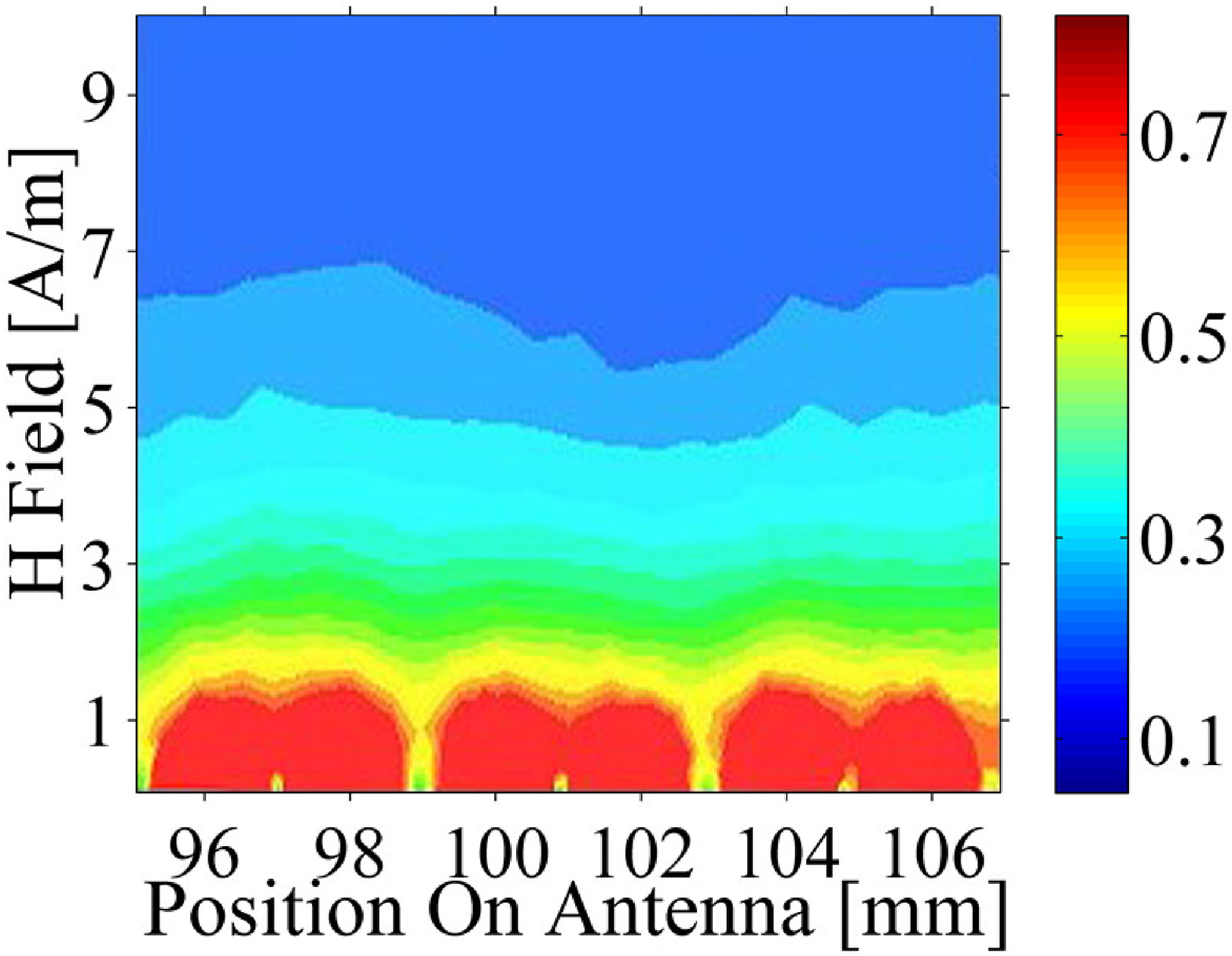}}
\subfigure[] {\includegraphics[width=0.24\textwidth]{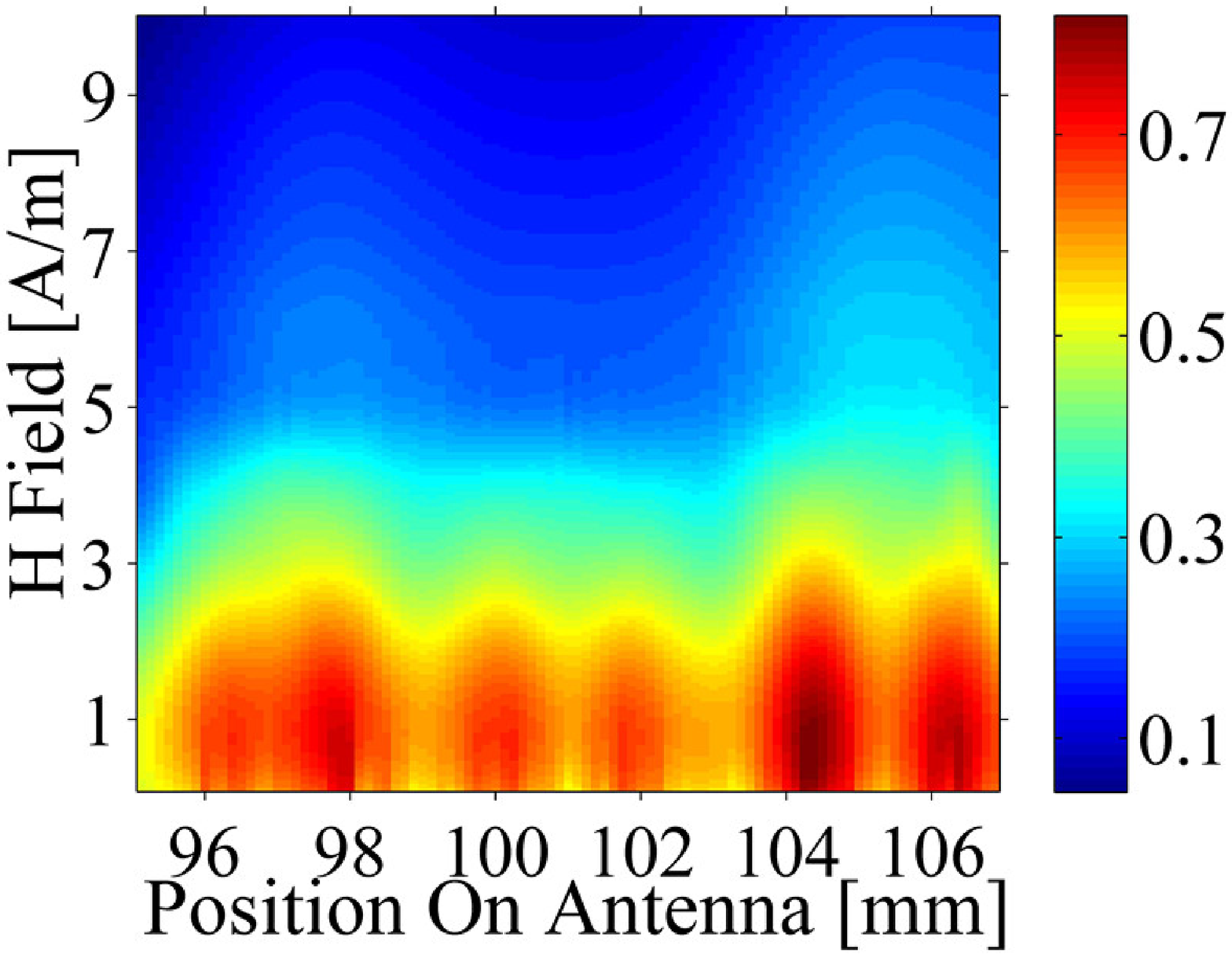}}
\caption{(a) The simulation result of the MW field distribution on the surface of antenna. The simulation tool is Ansoft HFSS, the mesh density is set to auto-adaptive. (b) The actually measured result on experimental system.  The MW imaging is measure by ODMR spectrum analysis method.}
\label{Fig11}
\end{figure}

For computing accurate MW field strength measurement results, we firstly use Rabi measurement sequence to control experimental system to acquire CCD imaging data, and then use matlab to fit measured data. As an example, we can see from Fig.\ref{Fig9}, the Rabi oscillations decay at a very fast rate, ${\tau _R}$ is 71 ns, which decrease the sensitivity of detection. The measurement result of Ramsey oscillation in experimental system refer to Fig.\ref{Fig10}. The calculation results show that the $T2$ of diamond sample is 52 ns.
ODMR measurement method can achieve higher signal-to-noise ratio, because of the time integration effect of CCD and the spatial accumulation effect of ensembles NV color centers. Owing to the short ${\tau _R}$, the sensitivity of Rabi frequency measurement is not satisfactory, so we use the area of resonance peak in ODMR spectrum to calculate the strength of MW near field. Compared with other resonance points in ODMR spectrum, the resonance point whose frequency is same to antenna working frequency provide perfect SNR than other resonance points. In order to verify the correctness of measurement results, we used Ansoft HFSS 15.0 to simulate the near field distribution of antenna. Fig.\ref{Fig11}(a) shows the simulation result of HFSS.

The principle of HFSS is based on the FEM method of computational electromagnetics. Thus, the user must carefully design the size of mesh for balance between computational efficiency and complexity. Because of the mesh, the MW field images generated by HFSS software exhibit zigzag irregular patterns along the edges of the mesh. The direct MW imaging technology will free users from this issue. Fig.\ref{Fig11}(b) presents the MW near field on the surface of spiral antenna, which is measured by the experimental system. Owing to the limitation of the diamond material, we used the ODMR spectrum analysis algorithm to reconstruct the MW imaging.

In order to analyze the reason of short ${\tau _R}$ of diamond, we analyzed electron spin resonance (ESR) spectra of diamond samples. There are three electron spin structures in diamond sample, namely P1 center spin, board spin and narrow spin. P1 center spin is a kind of defect formed by the substitution of one nitrogen atom for one carbon atom. The P1 center included an unpaired electron, and the spin g factor was equal to 2.00276. The proportion of the P1 center was $24{\rm{\% }}$. Broad spin corresponded to the broken bond of carbon atoms on the surface of diamond particle. The broad spin included an unpaired electron, the measured spin g factor is 2.00372, and the proportion of broad spin is $63{\rm{\% }}$.
The narrow spin corresponded to the vacancy defect in the diamond, which also included an unpaired electron, the measured spin g factor was 2.00351, and the proportion of narrow spin was $3{\rm{\% }}$. We thus conclude that in the ESR spectrum analysis result, the broad spin is a mainly component in the diamond sample. With the diameter of diamond particles becoming smaller, the proportion of broad spin will become increasingly larger. Compared with P1 center, the percentage of narrow spin is obviously low. Therefore, even if the 800 degree annealing process is adopted, the proportion of NV color centers is still not high. This conclusion is consistent with our experimental test results. The high proportion of P1 center and broad spin is the reason of short ${\tau _R}$.

Compared with Fig.\ref{Fig11}(a), the spatial resolution of Fig.\ref{Fig11}(b) has been remarkably improved. Because the actual antenna is not an ideal geometry, so the near field distribution image of MW quantum can reflect the actual MW near field distribution of the real antenna more accurately. This technique makes it possible for researcher to visual design MW antennas and devices.

\section{CONCLUSION}
Visualization of the MW near field distribution at the chip scale has been a long standing demand for decades in the design and fabrication process of integrated MW circuits, in particular in recently years. However, there is no satisfactory method in terms of invasiveness, spatial resolution and distance to chip surface. This paper presented a quantum approach for imaging and quantitative measurement of MW near fields based on the diamond NV center. We implemented a pulsed scheme to synchronize the laser, MW pulse and the frame capturing of camera. Compared to the previously mentioned vapor cell, electro-optic and spin-tronic sensing, this quantum approach uses an ultrathin diamond film with micrometer in thinness, which can be brought to the very close proximity of a MMIC chip surface. Since the sensing elements are atomic defects inside the dielectric diamond film, the invasiveness is minimized. Furthermore, the field strength deduced from the quantum state evolution is naturally calibrated. This method presents a quantitative measurement of the field strength. The field vector can be fully reconstructed based on the four crystalline axes. The algorithm of which is also presented in this work. However, limited by the sample preparation, in this work, a diamond particle, instead of diamond thin film, with a micrometer size was used to demonstrate the field pattern of a spiral antenna. Experimental results were compared with HFSS simulation data, and good agreement is obtained. In the following research, we will prepare a diamond thin film containing NV center in the surface layer with a thinness of micrometers, and we will test this method on an MMIC chip to realize the chip scale fault diagnosis and failure analysis.

\section{Acknowledgment}
This work is supported by Jiangsu Distinguished Professor program (Grant No. RK002STP15001) and NJUPT Principal Distinguished Professor program (Grant No. NY214136). Bo Yang would like to acknowledge valuable guidance to HFSS simulation from Prof. Bo Li and Mr. Yixiong Zhang of College of Electronic and Optical Engineering \& College of Microelectronics, NJUPT. Supports on ESR experiments from Mr. Runxiang Zhou of Peking University is also acknowledged.

%

\ifCLASSOPTIONcaptionsoff
  \newpage
\fi




%

%

\begin{IEEEbiography}[{\includegraphics[width=1in,height=1.25in,clip,keepaspectratio]{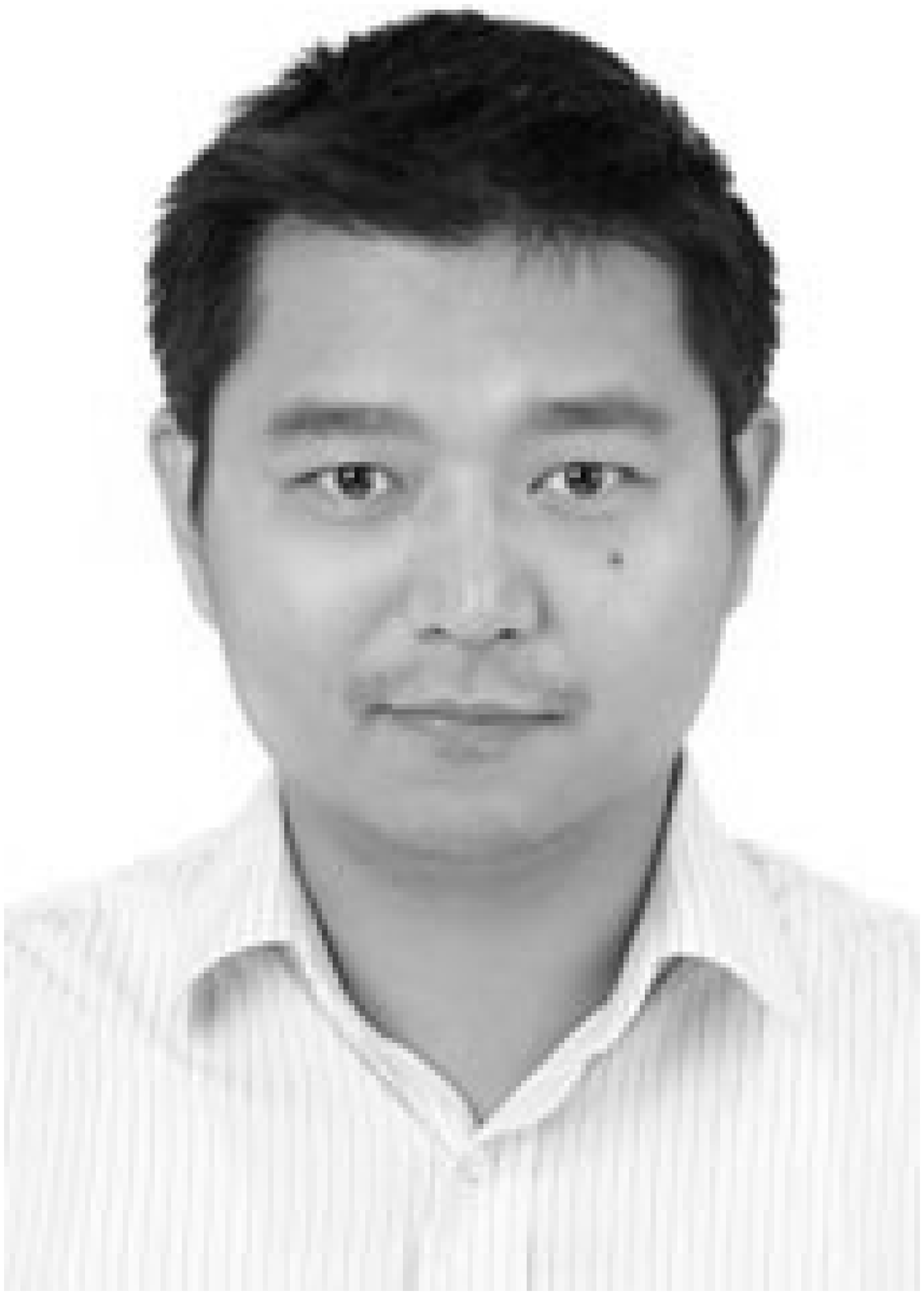}}]{Bo Yang}
was born in Nanyang district, Henan province, China, in 1980. He received the B.S. degree from Zhengzhou University in 2001, and Master degree from Xidian University in 2005. He has been studying for a doctorate at Nanjing University of Posts and Telecommunications since 2016. His research interests include quantum magnetic field imaging and measurement.

He worked with the ZTE  telecommunication cooperation from 2005 to 2013, at Nanjing research center, engaged in the development of 3G/4G mobile network. During this period, he was responsible for technical support of several international projects, such as H3G Austria and Deutsche Telecom German. After 2013, He joined National RFID Production Quality Supervision and Testing Center of china, the major field is electromagnetic compatibility testing.
\end{IEEEbiography}
\begin{IEEEbiography}[{\includegraphics[width=1in,height=1.25in,clip,keepaspectratio]{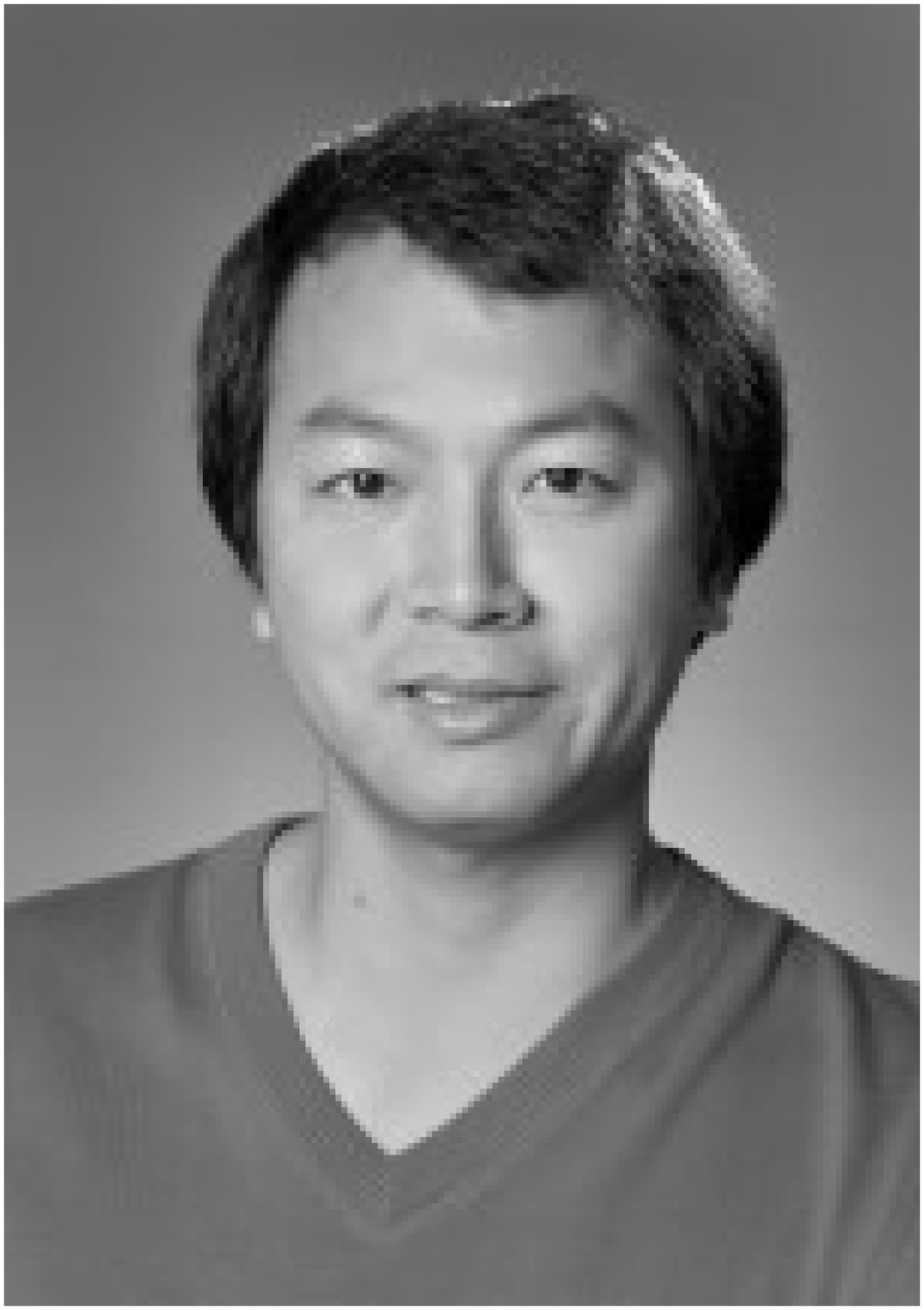}}]{Guan-Xiang Du}
received the B.S. degree from Lanzhou University in 2003, and the Ph.D. degree in physics from the Institute of Physics, Chinese Academy of Sciences,in 2008.

From 2008 to 2012, He was a postdoctor researcher in Takahashi Laboratory, Tohoku University, afterwards he joined Quantum Atom Optics Group, University of Basel. From 2014 to 2016, he worked with Nanoscale Spin Imaging Group, Max Planck Institute for Biophysical Chemistry.
He joined Nanjing University of Posts and Telecommunications, China, in 2016, as a professor. His research interests include spintronics, instrumentation on magneto-optical spectroscopy, nanoscale magnetometry and quantum MW field imaging.
\end{IEEEbiography}
\begin{IEEEbiography}[{\includegraphics[width=1in,height=1.25in,clip,keepaspectratio]{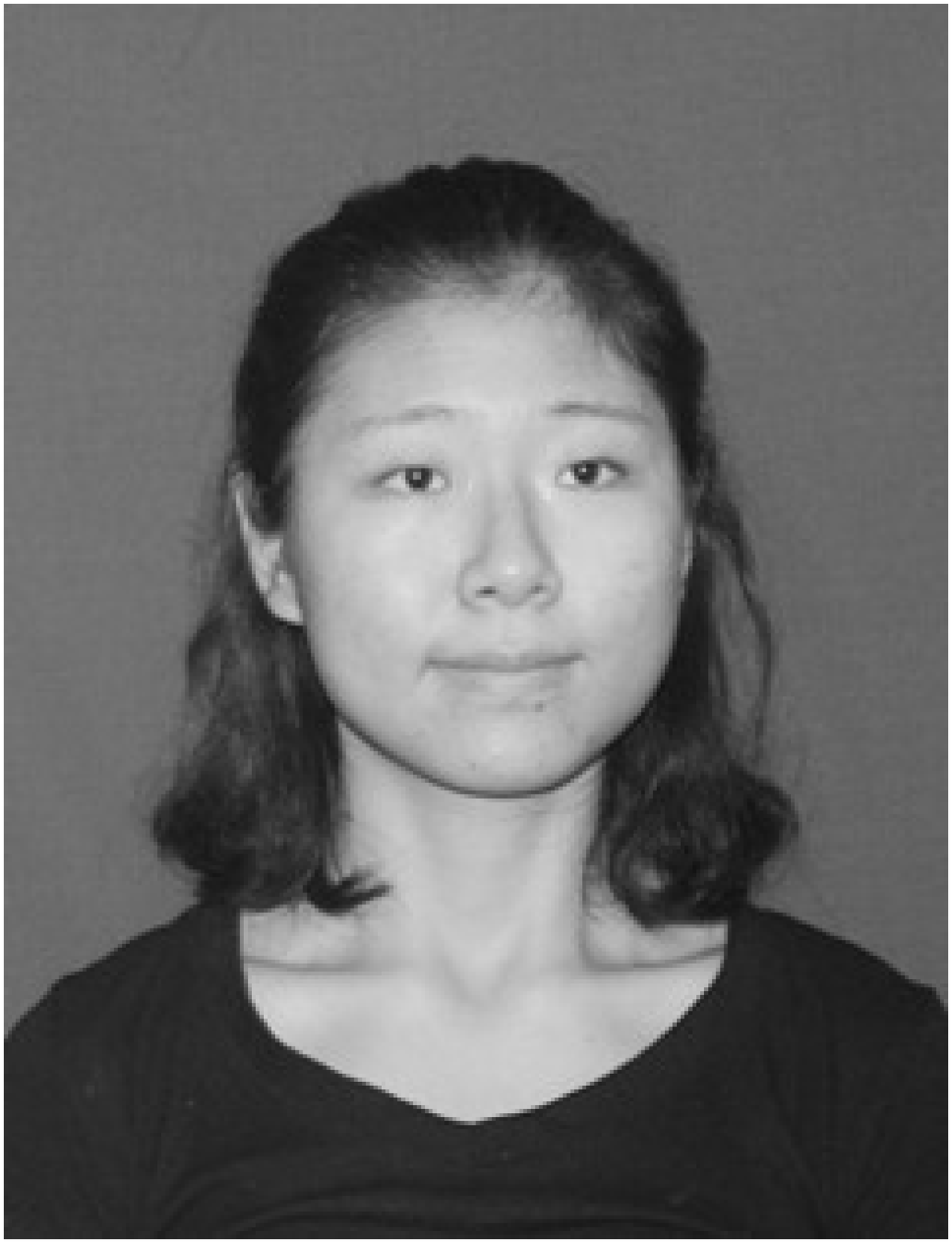}}]{Yue Dong}
was born in Zhenjiang, Jiangsu province, China. She is studying for a bachelor¡¯s degree at Nanjing University of Posts and Telecommunications from 2014 and majored in communication engineering.

From 2014 to 2017, she is a student in Nanjing University of Posts and Telecommunications. Yue Dong participates in the research project on the imaging and measurement of microwave near field based on diamond sensor.
\end{IEEEbiography}
\begin{IEEEbiography}[{\includegraphics[width=1in,height=1.25in,clip,keepaspectratio]{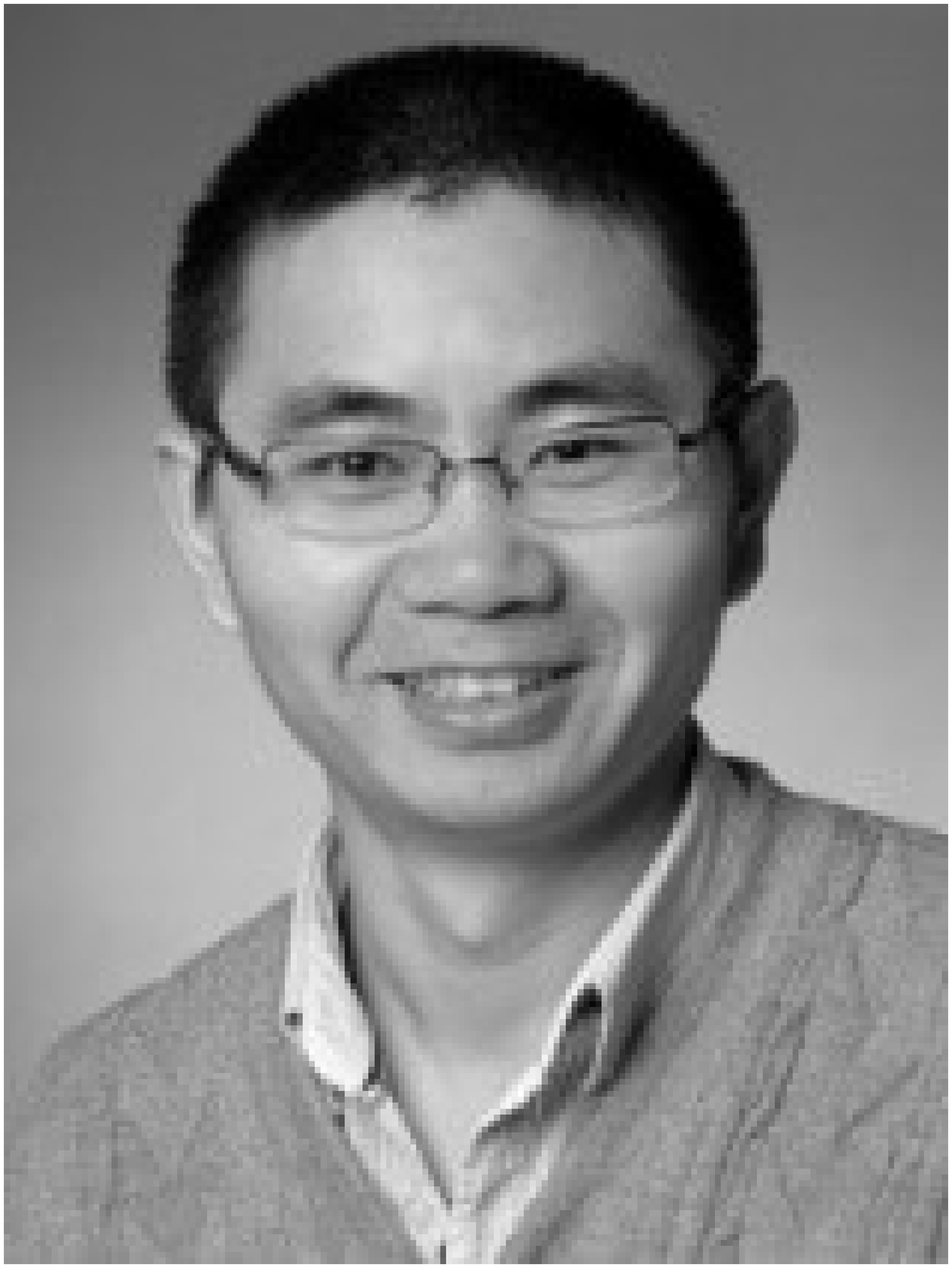}}]{Guo-Quan Liu}
received his B.S. degree in 2004 from Xiangtan University in China in 2004, and his Ph.D degree in 2012 from University of Oxford in UK. From 2012 to 2016, he worked as a postdoc researcher in the research group of Electron Spin Resonance in the Max Planck Institute for Biophysical Chemistry in Germany. From the beginning of 2014 he received a postdoc fellowship from the Humboldt Foundation. In 2016, he joined the Peking University as an assistant professor. His current research interests include methodological development to enhance the sensitivity of nuclear magnetic resonance and the biomedical applications of magnetic resonance.
\end{IEEEbiography}
\begin{IEEEbiography}[{\includegraphics[width=1in,height=1.25in,clip,keepaspectratio]{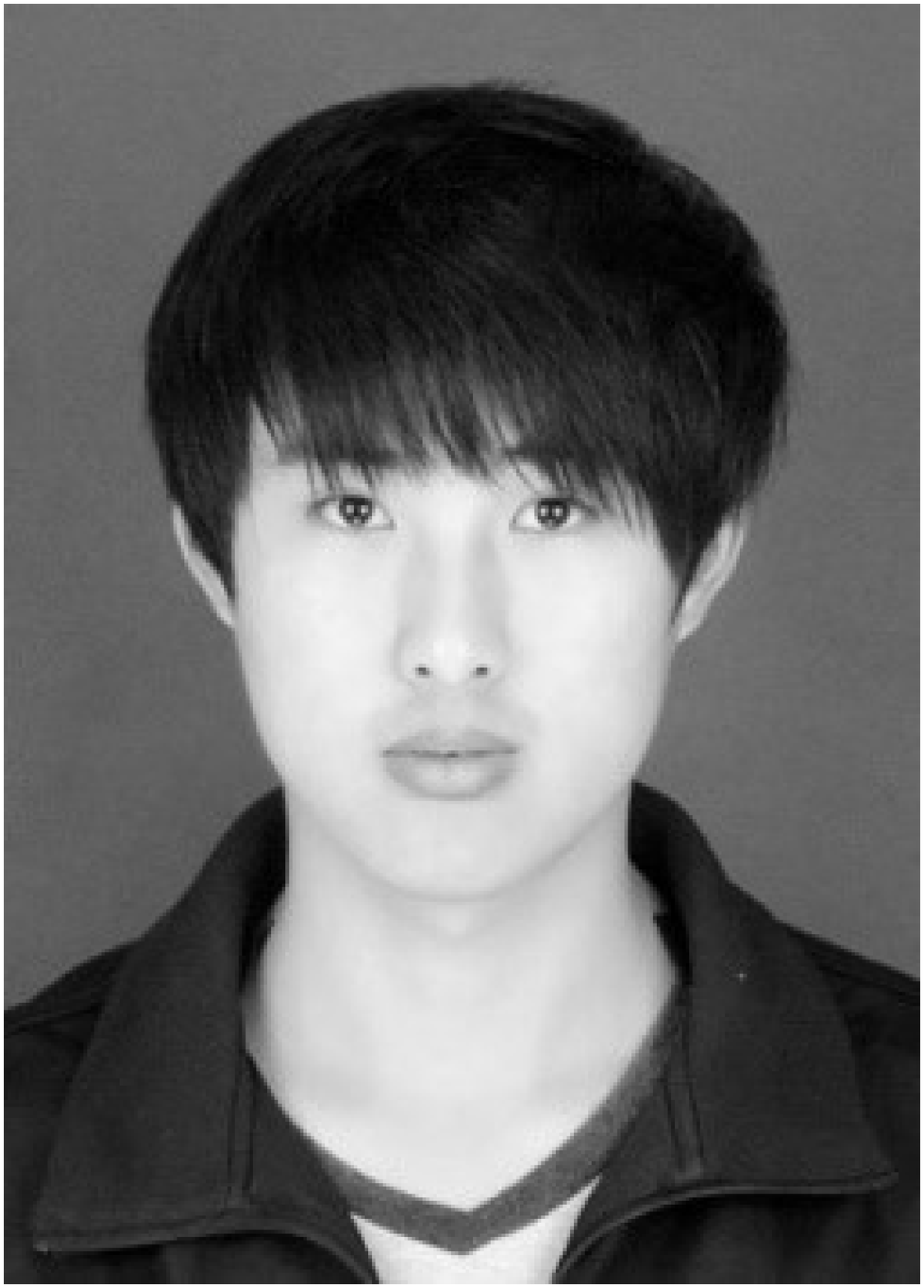}}]{Zhen-Zhong Hu}
was born in Xuzhou, Jiangsu province, China in 1993. He received the B.S. degree in communication engineering from Nantong University in 2016, and he is studying for a Master degree in College of Telecommunication \& Information Engineering at Nanjing University of Posts and Telecommunications from 2016. His research interests include quantum MW field imaging and detection.
\end{IEEEbiography}
\begin{IEEEbiography}[{\includegraphics[width=1in,height=1.25in,clip,keepaspectratio]{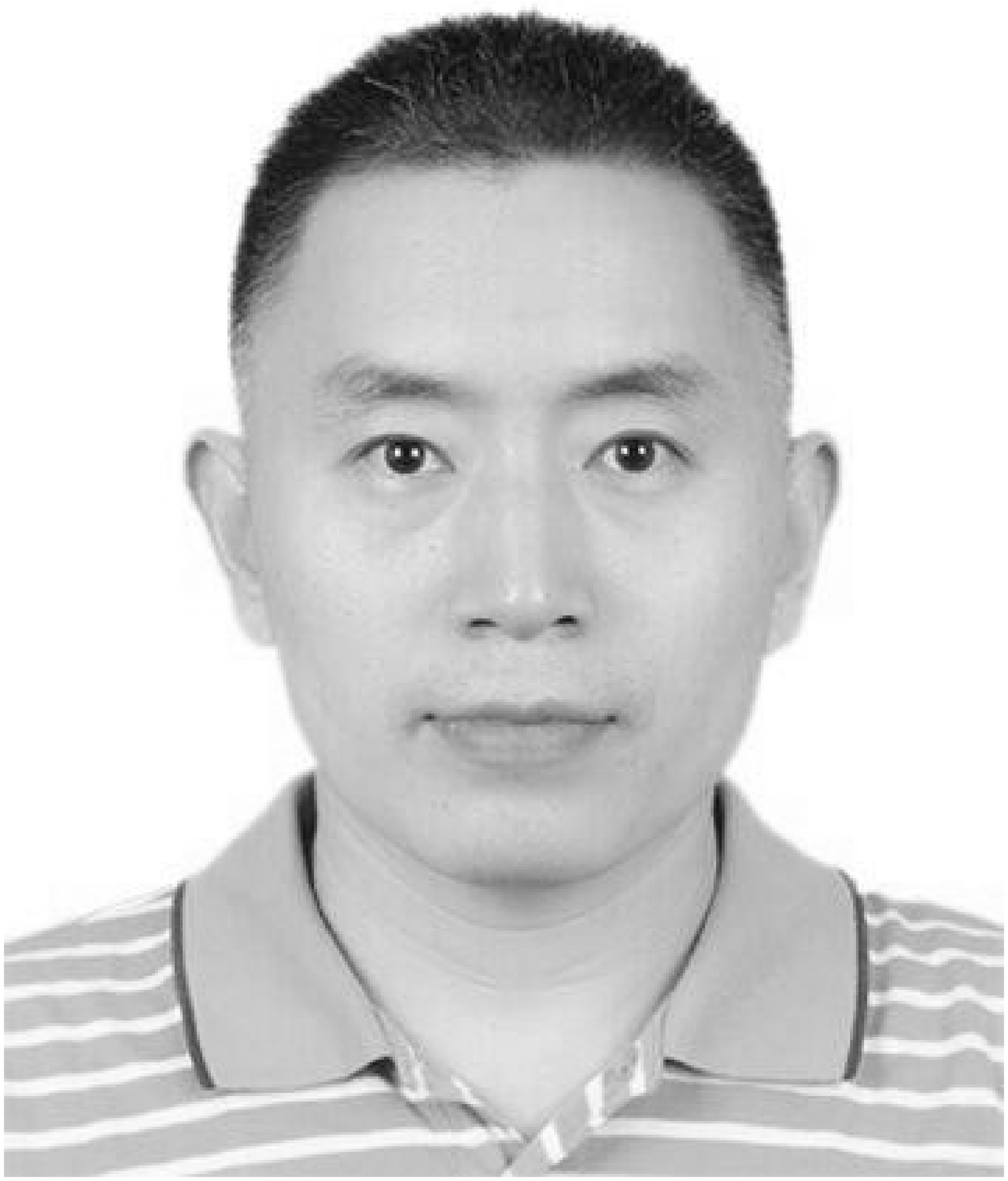}}]{Yong-Jin Wang}
received his PhD at Shanghai Institute of Microsystem and Information Technology, Chinese Academy of Sciences in 2005. He held various research positions in Germany, Japan, and UK, respectively. Since 2011, he has been a professor at Nanjing University of Posts and Telecommunications. Now, he is the chief investigator of national innovation base for micro-nano device \& information system. His current research is to conduct III-nitride monolithic photonic circuit for visible light communication and the Internet of Things.
\end{IEEEbiography}



\end{document}